\documentclass[fleqn,twoside]{article}
\usepackage{espcrc2}

\usepackage{graphicx}
\usepackage[figuresright]{rotating}

\newcommand{\AmS}{{\protect\the\textfont2
  A\kern-.1667em\lower.5ex\hbox{M}\kern-.125emS}}

\title{Experimental study of quasi-elastic neutrino interactions on
Ar with a liquid Ar TPC exposed to the WANF neutrino beam}

\author{A. Curioni\address[Yale]{Physics Department, PO BOX 208120, Yale University,
        New Haven, CT, 06520-8120, USA } \thanks{On behalf of the ICARUS-Milano
        collaboration}}
       
\begin{document}

\begin{abstract}
We present results from the first exposure of a liquid Ar time
projection chamber to a neutrino beam. The data have been collected in
1997 with a 50 liter ICARUS-like chamber located between the CHORUS and
NOMAD experiment at the CERN West Area Neutrino Facility. We focus on
the analysis of quasi-elastic interactions; despite the limited size of
the detector, nuclear effects beyond Fermi motion and Pauli blocking
have been observed as perturbations to the pure quasi-elastic
kinematics.
\vspace{1pc}
\end{abstract}

\maketitle

\section{Introduction}

In 1997 a liquid argon time projection chamber (LArTPC) with a fiducial
volume of about 50~l was exposed to the multi-GeV wide-band neutrino
beam of the CERN West Area Neutrino Facility
(WANF)~\cite{ref:wanf,ref:flux_nomad}, during the NOMAD~\cite{NOMAD}
and CHORUS~\cite{CHORUS} data taking. 
The LArTPC was placed on a platform 4.5~m high, between the CHORUS
and NOMAD detectors. The modest size of the LArTPC fiducial volume
($\sim$50 liters) made necessary a muon spectrometer downstream the
TPC. A coincidence with the NOMAD DAQ was set up to use the detectors
located into the NOMAD magnetic dipole as a spectrometer.
During the test $1.2 \times 10^{19}$ p.o.t. were integrated, and about
$10^5$ triggers recorded. The test was run jointly by a group from
Milano University and the ICARUS collaboration~\cite{Stefano}. 
The data accumulated in 1997 still offer the unique opportunity of an
experimental study on neutrino interactions on Ar nuclei.
At this time of writing, final results from the experiment have not
yet been reported in any systematic fashion (but see
\cite{ref:pietropaolo,ref:tesi} for preliminary results and a more
detailed description of the experimental apparatus); a comprehensive
set of results from the 1997 test is presented here for the first
time, while a more thorough and final paper is in preparation. 

\section{The experimental setup}
\label{Sec:Setup}

The LArTPC had an active volume of 32$\times$32$\times$46.8~cm$^3$,
enclosed in a stainless steel vessel in the shape of a bowed-bottom
cylinder 90 cm high with a radius of 35 cm. The active volume
contained 67 kg of liquid Ar. Ionization electrons produced by
the passage of charged particles in liquid Ar drift toward the anode
under the action of 214~V/cm uniform electric field. The readout
electrodes are two parallel planes of wires running orthogonally; the
planes are at a distance of 4~mm and each plane has  128 wires. Each
stainless steel wire has a diameter of 100~$\mu$m; the distance
between the wires is 2.54~mm. 
The ionization electrons drift through the first ({\em induction})
wire plane so that the integrated induction signal is zero.  
The {\em collection} plane then collects the drifting electrons. The
mean charge per unit  pitch (2.54~mm) for a m.i.p. crossing the
chamber parallel to the readout planes, corresponds to about $2.3
\times 10^4$ electrons ($\sim$4 fC).

The chamber has been exposed to the $\nu$ beam produced at the CERN
WANF. The
primary 450~GeV protons from the CERN SPS were extracted every 14.4~s
in two spills  ($1.8 \times 10^{13}$ protons per spill, on average) of
6~ms duration each and separated by 2.5~s, and hit a segmented Be
target. Secondaries were selected in momentum and focused by a system
of collimators and magnetic lenses. 
The experimental area was located 940~m downstream the target. The
mean energy of the $\nu_\mu$ reaching the detectors is 24.3~GeV, while
contaminations from other flavors are below 7\% for $\bar{\nu}_\mu$
and $\sim 1$\% for $\nu_e$~\cite{ref:flux_nomad}.

The trigger was provided by a set of plastic scintillators, located
downstream the chamber before the NOMAD apparatus. 
Incoming charged particle were vetoed by 5 large plastic scintillators
mounted in front of the chamber and by the last scintillator plane of
CHORUS, which vetoed particles deflected by the CHORUS magnetic field
entering the chamber at large angles with respect to the nominal beam
direction.   

The local trigger required the coincidence of the SPS beam spill, at 
least one of the trigger scintillators, and the two trigger
scintillator planes of NOMAD (T1 and T2 in \cite{ref:trigger_NOMAD}). 
Moreover, a trigger was rejected if the NOMAD acquisition system was
in \verb+BUSY+ mode or if it came less than 500~$\mu$s after the
previous trigger, inhibiting the occurrence of overlapping. In this 
configuration $\sim 11$\% of the recorded events had a vertex in the
fiducial volume of the TPC. 
The request of a charged particle triggering the chamber locally and
reaching NOMAD up to T1 and T2 limited the useful data to $\nu_\mu$
charged-current interactions.  
  
\section{Event reconstruction and particle identification}
\label{Sec:event_reconstruction}

\begin{figure}[t]
  \includegraphics[scale=0.7,bb=205 210 470 585,clip=true]{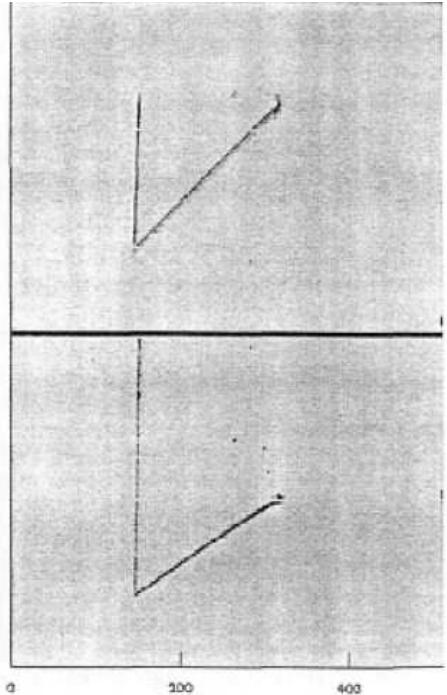}
  \caption{The filtered image of an event reconstructed as
  $\nu_{\mu} n \rightarrow \mu^- p$, with a m.i.p. leaving the TPC
  fiducial volume and an identified stopping proton. {\em Top}
  induction view; {\em bottom} collection view.}
  \label{fig:evt}
\end{figure}

Interactions in the TPC fiducial volume were fully imaged in two 2D
images with a common coordinate (time), with full calorimetric
information associated with each point. The common coordinate allows
an unambiguous 3D reconstruction of the event, at least for simple
topologies. 
Each 2D-image represents the signal amplitude digitized by
the ADC (in a linear gray scale) versus the time sample (drift
coordinate) and the wire number.  Fig.~\ref{fig:evt} shows the
image of a quasi-elastic (QE) event in the induction and collection 
planes. Units are wire number versus drift time in $\mu$s.   
Appropriate filtering techniques have been applied to deconvolve the
instrumental response and precisely associate the charge deposition to
each pixel in the image.   
For the collection wires a universal response function has been
determined empirically from events in the m.i.p. sample (elementary
charge deposition).  
The response was then deconvolved by means of a Discrete Fourier
Analysis to reconstruct the topology of the actual charge deposition
in the event.  
A different approach was adopted for the induction view, where the
charge measurement in that view was not accomplished and only the
event position, measured by the zero-crossing of the induced signal,
was reconstructed. 

\subsection{Proton reconstruction and p/$\pi$ separation}
\label{sec:p_pi}

For a proton, the identification and momentum measurement were
performed using only information from the TPC. The discrimination
between protons and charged pions is performed exploiting the
different behavior of energy loss as a function of the range. For
the case of candidates stopping in the fiducial volume of the chamber, 
as the ones of the ``golden sub-sample'' considered in
Sec.\ref{Sec:Analysis_QE}, the discrimination is highly simplified: it
is based on the consistency between the reconstructed kinetic energy
from the range and the integrated charge deposited by the candidate
along its path-length. 
For partially contained candidates, a much better assessment of the
charge response of the TPC and of quenching effects\cite{TMG} is
required, since proton identification is based only on the visible
pattern of the energy loss. Such events are not considered here. 
For fully contained protons, the momentum uncertainty is dominated by
the finite pitch of the wires. The equivalent pitch in the vertical
direction (drift direction) is much smaller due to the high sampling
rate of the fast ADC ($360\ \mu$m). For horizontal protons of
$p=400$~MeV/c the uncertainty is of the order of 7~MeV/c. The angular
resolution in the collection or induction view depends on the number
of wires $N$ hit by the particle along its path; it is $\sigma \simeq
0.36\sqrt{12}/(2.54\ N^{3/2})$, corresponding e.g. to 15 mrad for
N=10.  

\subsection{Muon reconstruction}
\label{ref:muon_rec}

The kinematic reconstruction of the outgoing muons exploits the
tracking capability of NOMAD. An event triggering the chamber will
have at least one penetrating track reaching the T1 and T2 trigger
scintillators bracketing the TRD of
NOMAD~\cite{ref:trigger_NOMAD}. The corresponding track, nearly 
horizontal at the entrance of the NOMAD drift chamber volume, is
reconstructed with an average momentum precision of $\sigma_p/p \sim
0.05/\sqrt{L} \oplus 0.008p/L^{5/2}$, $L$ being the visible range in
the volume itself expressed in meters and $p$ the particle momentum in
GeV. 
A 10~GeV horizontal muon crossing all the chambers ($L\sim 5m$) is
reconstructed with a precision of 2.2\%.  The reconstructed particle
is traced back to the TPC accounting for the magnetic field and the
presence of the forward NOMAD calorimeter. The latter introduces the 
dominant uncertainty on the muon transverse momentum, due to multiple
scattering (MS) in iron (190~cm for a horizontal muon). For small
scattering angles ($\theta \ll 1$~rad) the MS uncertainty on the
transverse momentum is independent of $p$ and it turns out to be $\sim
140$~MeV. The correctness of the back-tracing procedure has been
cross-checked comparing the direction angles of the particles
belonging to the m.i.p. calibration sample as measured by the TPC with
the corresponding quantity from NOMAD. 


\section{Analysis of quasi elastic $\nu_\mu$ interaction} 
\label{Sec:Analysis_QE}

\subsection{Event selection and rates}
\label{Sec:Sel_QE}

The commissioning of the detector ended in August 1997. After that,
$1.21\times 10^{19}$ protons on target have been integrated. The
trigger efficiency was monitored during data taking and its integrated
value is 97\%. Additional losses are due to the TPC (3\%) and NOMAD
(15\%) dead time and to detector faults. The effective livetime was
75\%. 81,000 events had been recorded and analyzed by visual
scanning. A minimum-bias sample has been obtained requiring no more
than two tracks exiting the chamber and an arbitrary number of
contained tracks. This sample contains the {\em golden sample} of
QE interactions and other samples used to validate the simulation for 
background estimation. The golden sample consists of events with an
identified proton of kinetic energy ($T_p$) larger than 50~MeV fully
contained in the TPC and one muon whose direction extrapolated from
NOMAD matches the outgoing track in the TPC.  
The distance of the interaction vertex from any of the TPC walls has
to be greater than 1~cm. The muon candidate track projected onto the
wire plane must be longer than 12 wire pitches. The event is accepted
even in the presence of other stopping particles, as far as their
kinetic energy (in the proton hypothesis) does not exceed the $T_p$ of
the leading proton. If tracks other than the identified muon leave the
TPC or at least one converted photon with energy greater than 10~MeV
is present in the fiducial volume, the event is rejected.  
The tightness of these selections defines a very clear topology for
visual scanning (Fig.~\ref{fig:evt}). 

The lower $T_p$ cut combined with the request of containment is very
severe since the chamber volume is small.
On the other hand, at lower $T_p$ the proton range is comparable with
the wire pitch and neither the proton momentum nor the interaction
vertex can be reconstructed with due precision.  Moreover, for
$T_p>50$ MeV the $\pi^{\pm}$/p misidentification probability is
negligible. The golden sub-sample contains pure QE interactions
($\nu_\mu \ n \ \rightarrow \mu^- \ p$), an intrinsic background
dominated by resonant productions followed by pion absorption in the
nucleus ( $\nu_\mu \ p \ \rightarrow \Delta^{++} \mu^- \rightarrow
\mu^- \ p \ \pi^+$ ,  $\nu_\mu \ n \ \rightarrow \Delta^{+} \mu^-
\rightarrow \mu^- \ p \ \pi^0$) and an instrumental background due to 
unidentified $\pi^0$'s ($\nu_\mu \ n \ \rightarrow \Delta^{+} \mu^-
\rightarrow \mu^- \ p \ \pi^0$). On the other hand, the $\nu_\mu \ n \
\rightarrow \Delta^{+} \mu^- \rightarrow \mu^- \ n \ \pi^+$
contamination is negligible in this tightly selected sample.

The efficiency of the selections for QE interactions and their
intrinsic background has been evaluated by Monte Carlo experimentation
based on the FLUKA code~\cite{ref:FLUKA} and it turned out to be 17\%.
Similarly, MC provided the inefficiency of the vetoing selections for
$\nu_\mu \ n \ \rightarrow \Delta^{+} \mu^- \rightarrow \mu^- \ p \
\pi^0$, i.e. the probability to miss both the decay photons of the
$\pi^0 \rightarrow \gamma \gamma$ or the $e^+e^- \gamma$ system in
case of $\pi^0$ Dalitz decay. The corresponding contamination of the
golden sub-sample is estimated to be 13\%, which can be checked by
test samples extracted from the minimum bias events. The test samples
consist of golden events with one ($N_1$) or two ($N_2$) converted
photons pointing to the interaction vertex. Assuming the gamma
identification probability to be uncorrelated for the two photons, we
have $N_1=2N(1-\epsilon_\gamma)\epsilon_\gamma$ and
$N_2=N\epsilon_\gamma^2$, $N$ being the (unknown) overall rate of $p \
\mu^- \ \pi^0$ final states and $\epsilon_\gamma$ the photon
identification efficiency; $\epsilon_\gamma = (N_1/2N_2 +1)^{-1}$
turned out to be (43$\pm$ 9) \%. 
Hence, the probability of missing both gammas is (32$\pm$10) \% to be
compared with the MC calculation of 20.4\%.  
Accounting for the detector livetime, trigger and selection efficiency
and the sample purity, we expect 73.5 events and we observe 61.

\subsection{Analysis of the kinematic distortion due to nuclear matter}

\begin{figure}
  \includegraphics[scale=0.6]{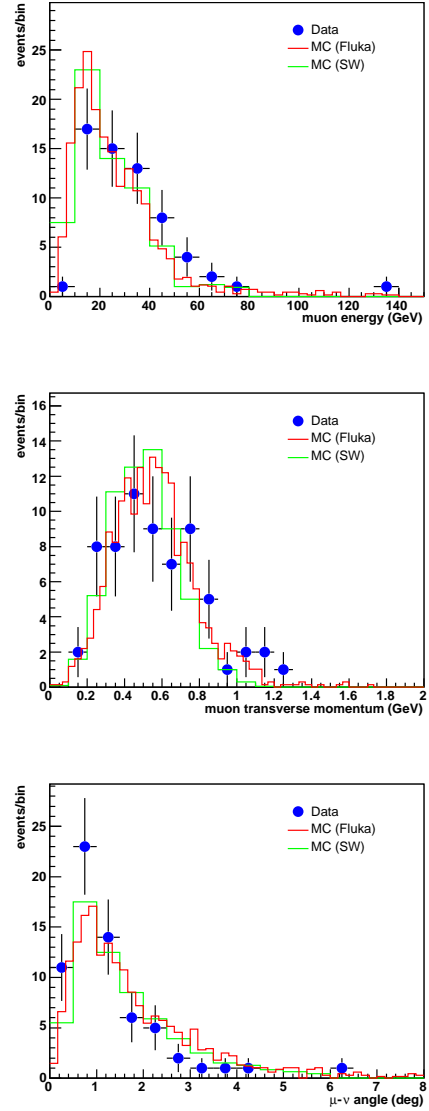}
  \caption{From top to bottom: distribution of the muon energy, muon
  transverse momentum and $\mu - \nu$ angle for the golden
  sub-sample. The continuous red (green) line is the expectation from
  FLUKA (Saxon-Woods) convoluted with the detector response.}
  \label{fig:1}
\end{figure}

\begin{figure}
  \includegraphics[scale=0.6]{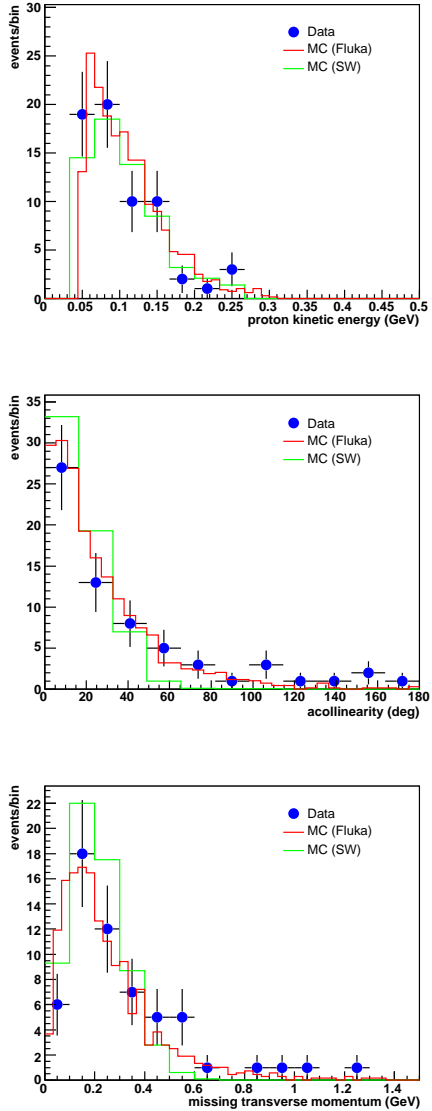}
  \caption{From top to bottom: proton kinetic energy, acollinearity
  and missing transverse momentum for the golden sub-sample. The
  continuous red (green) line is the expectation from FLUKA
  (Saxon-Woods) convoluted with the detector response.} 
  \label{fig:2}
\end{figure}

In spite of the limited statistics, the golden sub-sample provides
informations on the basic mechanisms that modify the kinematics of
neutrino-nucleus interactions with respect to the corresponding
neutrino-nucleon process. Nuclear matter perturbs the initial state of
the interaction through Fermi motion; it also affects the formation of
the asymptotic states through nuclear evaporation, hadronic
re-scattering or hadronic re-absorption.  Several kinematic variables
are only marginally affected by nuclear matters; in this case, the
corresponding distributions can be reproduced once the $\nu$-nucleon
interaction is corrected for Fermi motion and Pauli blocking using
e.g. a Saxon-Woods (SW) potential for the nucleus\footnote{In this
model, nucleons are treated as a Fermi degenerate gas located in a
potential well of the form $E(r) = E_0/(1+e^{(r-R)/a}) $ where $a$ is
0.6~fm; $E_0=46$~MeV is the maximum well depth and $R$ is the nuclear
radius ($R=3.6$~fm for Ar)~\cite{ref:balzarotti}.}. Clearly,
purely leptonic variables belong to this category.  
Fig.~\ref{fig:1} shows the distributions of muon energy, muon
transverse momentum and the $\mu - \nu $ angle. Deviations from
expectations are visible only for very low $\mu$ momenta and the
corresponding large scattering angles. $T_p$ (Fig.~\ref{fig:2}, top
panel) is, in principle, strongly influenced by the presence of
nuclear matter, but, in practice, the shape of the distribution is
determined by the $T_p>50$~MeV cut and the requirement of full
containment in the fiducial volume.    
Figs.~\ref{fig:1} and \ref{fig:2} (top panel) provide a useful
consistency check, and demonstrate that MC reproduces the kinematic
selection performed during the scanning and analysis of the golden
sub-sample. 

Other variables embedding the reconstructed kinematics of the protons
are sensitive to genuine nuclear effects and should depart from the
naive Saxon-Woods description. In particular, we analyzed both the
acollinearity and the missing transverse momentum of the event
(Fig.~\ref{fig:2}, middle and bottom panels). The former is defined as 
\begin{equation}
A \equiv \mathrm{acos} \left[ \frac{ p_{xp} \ p_{x\mu} + p_{yp} \ p_{y\mu} }
{\sqrt { (p_{xp}^2 + p_{yp}^2) (p_{x\mu}^2 + p_{y\mu}^2) } } \right]
\end{equation}
$p_{xp}$ and $p_{yp}$ being the transverse momentum components of the
proton and $p_{x\mu}$ and $p_{y\mu}$ the corresponding quantities for
the muon. For pure QE scattering on a nucleon the muon and the
proton has to be back-to-back in the transverse plane so that the
acollinearity is zero. 

\begin{figure}[tb]
  \includegraphics[scale=0.55, bb=10 1 272 272,clip=true]{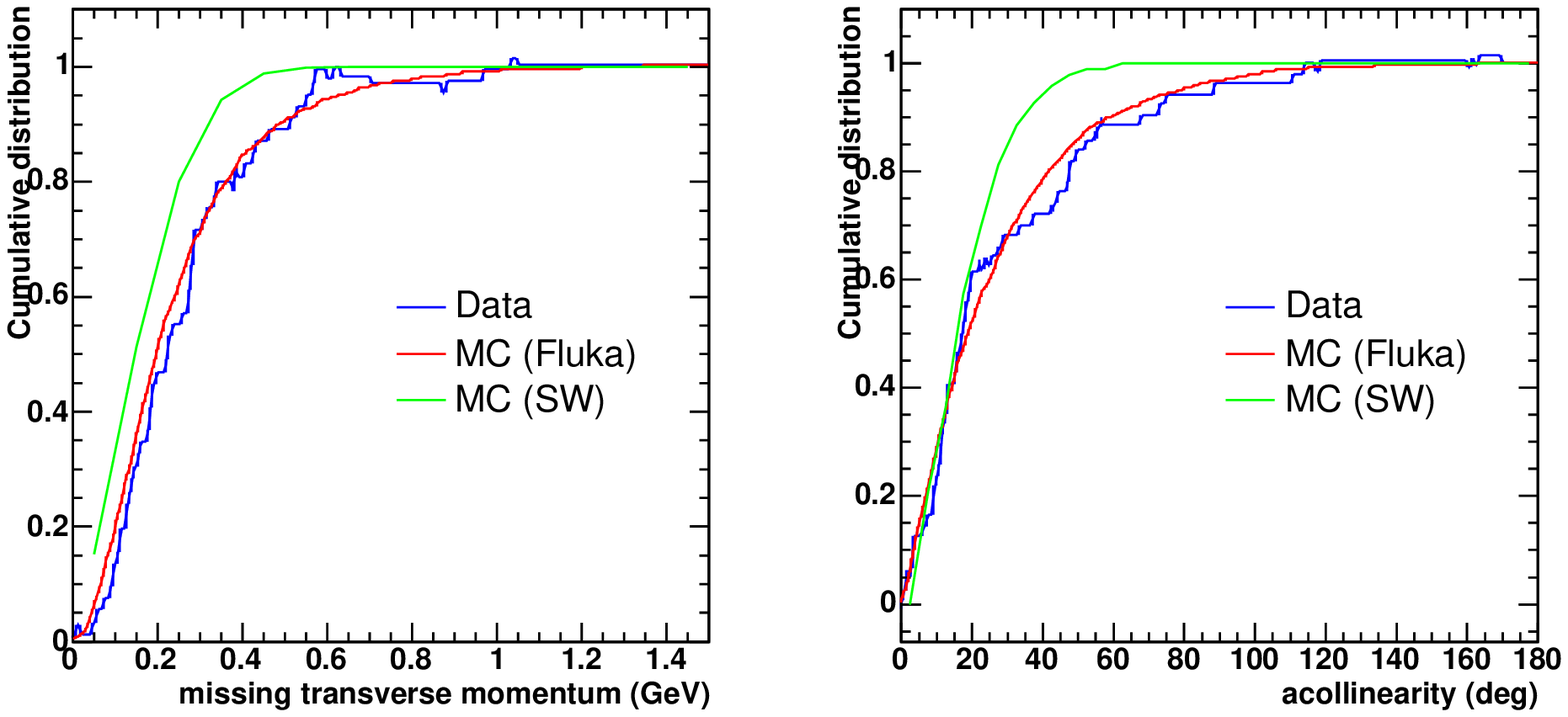}
  \includegraphics[scale=0.55, bb=293 1 560 272,clip=true]{cumulDist.eps}
  \caption{Background subtracted cumulative distribution of missing
  transverse momentum ({\em top}) and acollinearity ({\em bottom}).}
  \label{fig:3}
\end{figure}

In spite of the limited statistics and the $\sim10$\% contamination,
the imaging capability of the TPC allows to establish empirically the
inadequacy of the SW description of the nucleus. An improved
determination of the level of agreement between the theoretical
expectations and the data can be obtained from a background
subtraction procedure; this is based on the test samples mentioned in
Sec.\ref{Sec:Sel_QE}. If the QE analysis is applied to the test sample
where a $\pi^0$ is clearly identified, and the presence of the $\pi^0$
is then ignored, an artificial acollinearity excess is generated; in
particular, 36\% of the sample has $A>60^\circ$. Similar results are
obtained for $p_T^{\mathrm{miss}}$ where 72\% of the identified
resonance events shows a $p_T^{\mathrm{miss}}$ greater than
400~MeV. The corresponding background subtracted   are shown
in Fig.\ref{fig:3}.The Kolmogorov probabilities for acollinearity
(transverse momentum) are 0.52 (0.30) for Fluka and 0.003 (0.027) for
SW.  
%

\section{Conclusions}

We discussed the first exposure of a liquid Ar TPC to a multi-GeV
neutrino beam. The data provided relevant information to
experimentally establish the effectiveness of the LAr technology in
the reconstruction of low-multiplicity neutrino interactions. 
In spite of the limited size of the detector, nuclear effects beyond
Fermi motion and Pauli blocking  have been observed as perturbations
of the quasi-elastic $\nu_\mu$ CC interaction kinematics.

\section*{Acknowledgments}

We thank the NOMAD and CHORUS collaborations for their support during 
data taking and data analysis. The presenting author (A.C.) would like
to thank Bonnie T. Fleming of Yale.

\end{document}